\documentclass[12pt]{iopart}

\usepackage{graphicx}
\usepackage{amssymb}
\usepackage{lineno}

\newcommand{\bspec}{\mbox{$\beta$-spectrum}}
\newcommand{\belec}{\mbox{$\beta$-electron}}
\newcommand{\bdec}{\mbox{$\beta$-decay}}
\newcommand{\ev}{\mbox{\,eV/c$^2$}}
\newcommand{\evtwo}{\mbox{\,eV$^2$/c$^4$}}
\newcommand{\kr}{\mbox{$\rm ^{83m}Kr$}}

\begin{document}

\title[Precision HV divider, KATRIN experiment]{Precision high voltage divider for the KATRIN experiment}

\author{Th Th\"ummler$^{1}\footnote{Corresponding author. Present address: Institut f\"ur Kernphysik, Forschungszentrum Karlsruhe GmbH, Postfach 3640, 76021 Karlsruhe, Germany}$, R Marx$^2$ and Ch Weinheimer$^1$}

\address{$^1$ Institut f\"ur Kernphysik, Westf\"alische Wilhelms--Universit\"at M\"unster, Wilhelm-Klemm-Str. 9, 48149 M\"unster, Germany}
\address{$^2$ Physikalisch--Technische Bundesanstalt (PTB) Braunschweig, Bundesallee 100, 38116 Braunschweig, Germany}

\ead{thomas.thuemmler@ik.fzk.de}

\begin{abstract}
The Karlsruhe Tritium Neutrino Experiment (KATRIN) aims to determine the absolute mass of the electron antineutrino from a precise measurement of the tritium \bspec\ near its endpoint at 18.6~keV with a sensitivity of 0.2\ev .
KATRIN uses an electrostatic retardation spectrometer of MAC-E filter type for which it is crucial to monitor high voltages of up to 35~kV with a precision and long-term stability at the ppm level.
Since devices capable of this precision are not commercially available, a new high voltage divider for direct voltages of up to 35~kV has been designed, following the new concept of 
the standard divider for direct voltages of up to 100~kV developed at the Physikalisch-Technische Bundesanstalt (PTB)\footnote{The Physikalisch-Technische Bundesanstalt 
(PTB) is the German National Metrology Institute providing scientific and technical services.}.
The electrical and mechanical design of the divider, the screening procedure for the selection of the precision resistors, and the results of the investigation and calibration at PTB are reported here.
During the latter, uncertainties at the low ppm level have been deduced for the new divider, thus qualifying it for the precision measurements of the KATRIN experiment.
\end{abstract}

\maketitle

\section{Introduction}
\label{intro}

The properties of neutrinos and especially their rest mass play an important role for cosmology, particle physics, and astroparticle physics.
At present the most sensitive and  model-independent method to determine the neutrino rest mass in a laboratory experiment is the investigation of the energy spectrum of tritium \bdec .
Because of neutrino flavour mixing, the neutrino mass appears as an average of all neutrino mass eigenstates contributing to the electron neutrino.
At a few eV below the endpoint energy $E_0 = 18.6$ keV of the \bspec\ the signature of the neutrino rest mass is maximal \cite{otten_weinheimer2008}.
Until now only upper bounds on the neutrino mass of $m_\nu < 2$\ev\ have been determined \cite{pdg06,kraus_paper2005,lobashev_paper2003}.
In 2001 the international collaboration KATRIN \cite{KATRIN-LOI} was established to build a new tritium $\beta$-decay experiment.
The KATRIN experiment is based on the experimental experience of its predecessor experiments in Mainz \cite{kraus_paper2005} and 
Troitsk \cite{lobashev_paper2003} and aims to improve their sensitivity on the neutrino rest mass by one order of magnitude to $0.2$\ev\ \cite{KATRIN_design_report}.

KATRIN is using an integrating spectrometer of MAC-E filter type \cite{picard-mace, lobashev-85} for the energy analysis of the \bdec\ electrons.
For such a device the stability of the energy analysis relies primarily on the stability of the electrostatic filter potential \cite{otten_paper2006,kaspar}.
The stability of the latter has been identified as one of the five main contributions to the uncertainty of KATRIN \cite{KATRIN_design_report}.
In order to keep the high sensitivity on the neutrino mass, the contribution of the retarding potential to the systematic error has to be limited to $\Delta m^2 < 0.0075$\evtwo .
Any unknown filter potential fluctuation with a Gaussian variance $\sigma^2$ leads to a shift of the measured squared neutrino rest mass $\delta m^2_\nu$.
The general relation how systematic uncertainties affect the neutrino mass value is \cite{robertson}
\begin{equation}
\delta m_\nu^2 c^4 = - 2 q^2 \sigma ^2 \mathrm{,}
\end{equation}
with $q$ being the elementary charge.
Due to the given systematic error limit this leads to a maximum uncertainty of the filter potential of $\sigma < 0.061$\,V 
which corresponds to an allowed relative stability of the voltage monitoring system of $\frac{\Delta U}{U} < 3.3 \cdot 10^{-6}$ at a filter potential of $E_0/q = U = -18.6$\,kV.
This stability limit has to be kept for the anticipated KATRIN measurement time of three years, which corresponds to a calendar time of about five years.
The relative precision of the retarding potential is of primary importance, whereas the absolute precision, which includes also the contributions of work-functions and chemical shifts, is of less importance, since the endpoint of the $\beta$-spectrum is fitted from the data.
Still, the KATRIN experiment will use the absolute endpoint position obtained from the data to compare it to the  $^3$He-$^3$H mass difference \cite{nagy}, thus serving as an important check of the systematic corrections.\\
By precisely measuring the high voltage potential as well as by measuring the line position of a mono-energetic electron source, two redundant monitoring solutions will be applied \cite{KATRIN_design_report}. 
For the latter method the filter potential of the KATRIN main MAC-E filter will be continuously monitored by means of a monitor beam-line consisting of the MAC-E filter setup of the former Mainz neutrino mass experiment.
While being connected to the same filter potential as the main filter, mono-energetic electron calibration sources like \kr\ will be operated at the monitor beam-line.
Hence we are able to lock the main filter relative to the electron energy of the calibration source at the monitor filter.
For supplying the energy filters at both beam-lines with a stable high voltage potential we use state-of-the-art, commercially available, high voltage equipment with a stability of $5 \cdot 10^{-6}$ per eight hours.
High-end measurement equipment to monitor voltages in the 10 V range with a relative precision at the sub-ppm level is commercially available also.
What is not available is a precision high voltage divider to scale down the filter potential to the most sensitive 10 V range of the digital multimeter.
In \cite{marx}, a new concept of high voltage divider is reported, reaching $2 \cdot 10^{-6}$ per year relative stability for direct voltages of up to 100 kV.\\
The present paper describes how this concept has been applied to the new KATRIN precision high voltage divider for direct voltages of up to 35 kV\footnote{KATRIN is measuring \belec s; therefore, the filter potential has always a negative sign.
Since the divider operates at potentials of either sign, we omit the negative sign in its specification.
However, all measurements have been performed with negative voltages of up to -35 kV as required for calibration purposes when measuring all the conversion electron lines of \kr\ up to 32 keV.} using a different resistor technology and a simplified mechanical construction.
Finally, the new divider has been investigated and calibrated by comparison with the PTB standard divider of \cite{marx}.

\section{Design of the KATRIN Precision HV Divider}
\label{design}

\begin{figure}[t!]
\centering
\includegraphics[width=120mm]{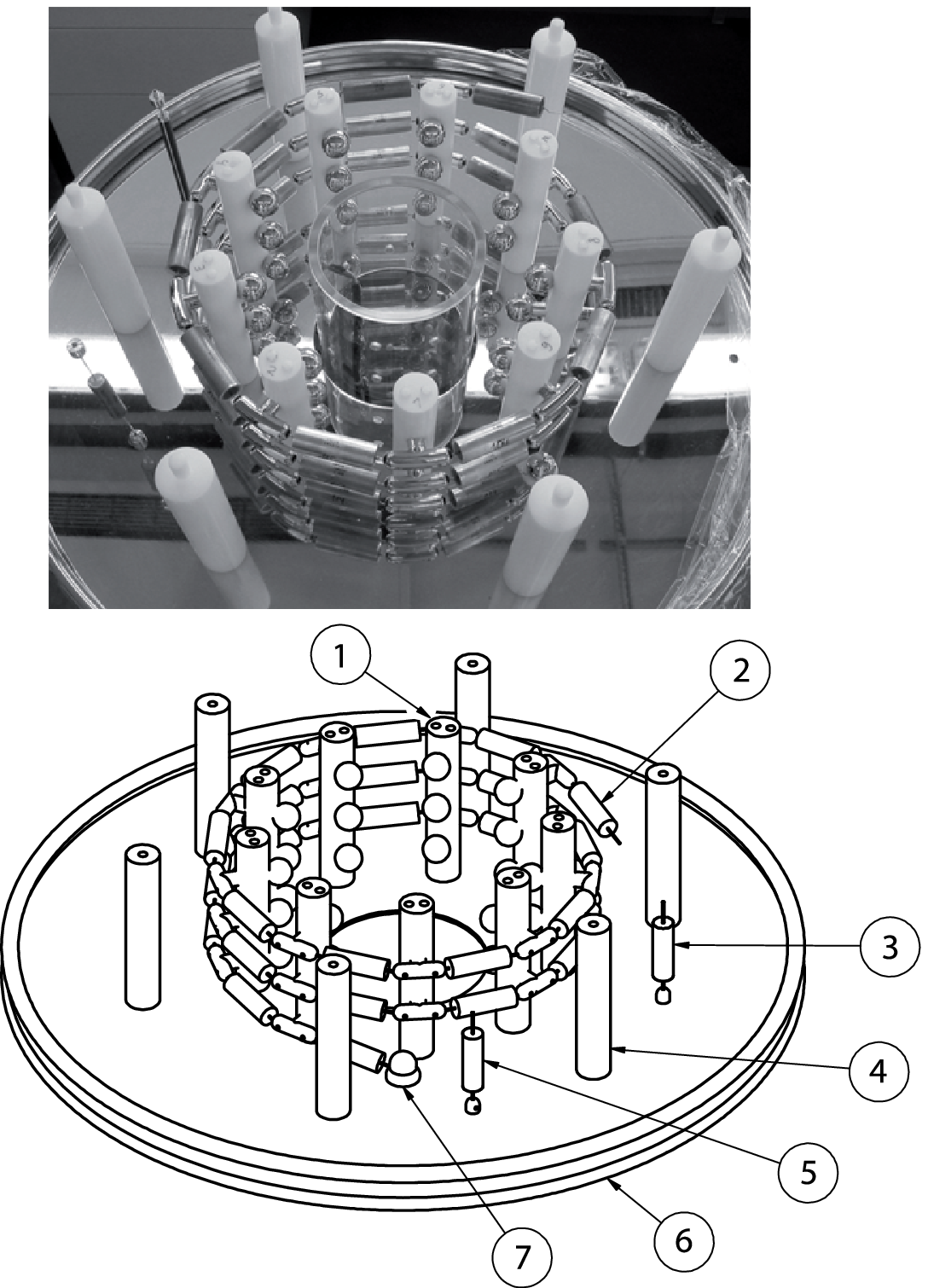}
\caption[Electrode layer assembly]{
Photo and layout of one electrode layer assembly. 1) PTFE precision resistor support. 2) Precision resistor of the high-precision divider. 3) HV resistor of the control divider. 4) POM rod electrode support. 5) HV capacitor of the control divider. 6) Copper electrode. 7) Insulated feedthrough of precision resistor chain to next layer.
}
\label{fig:electrode}
\end{figure}

\begin{figure}[t!]
\centering
\includegraphics[width=110mm]{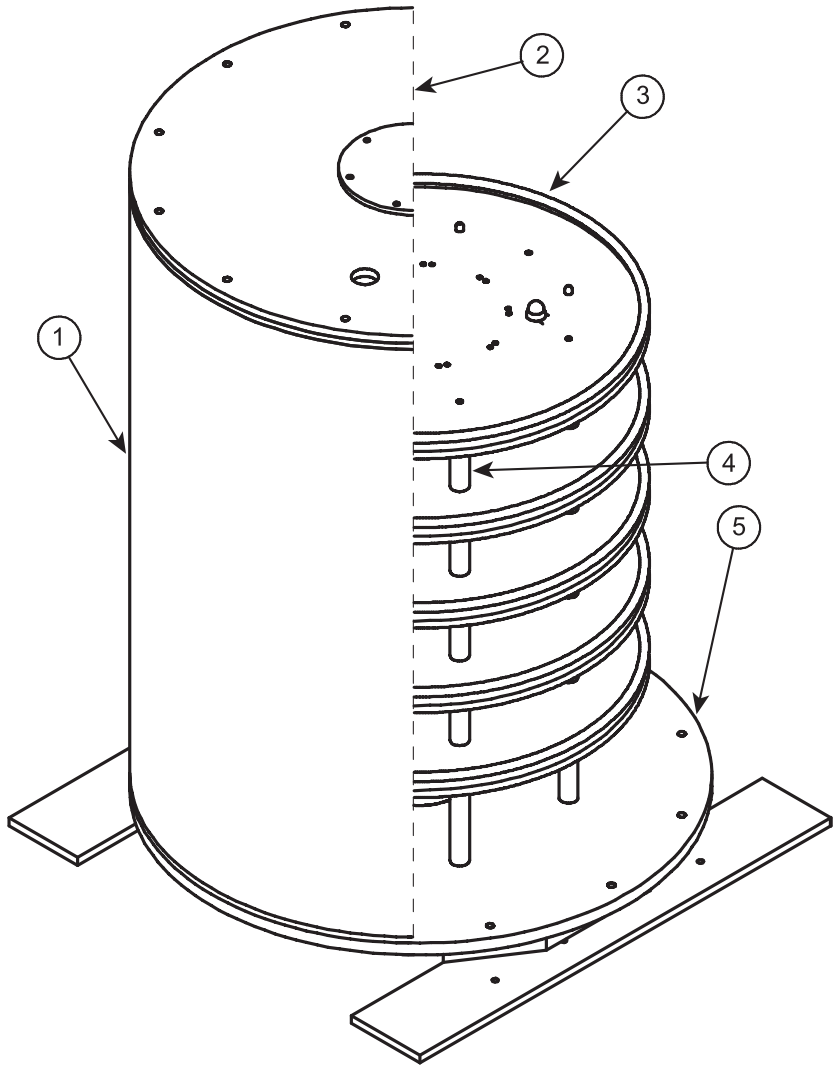}
\caption[Drawing of the divider setup]{
Drawing of the divider setup. 1) stainless steel vessel. 2) top flange. 3) control electrodes. 4) POM support structure. 5) bottom flange of the stainless steel vessel.
}
\label{fig:skizze1}
\end{figure}

\begin{figure}[t!]
\centering
\includegraphics[width=150mm]{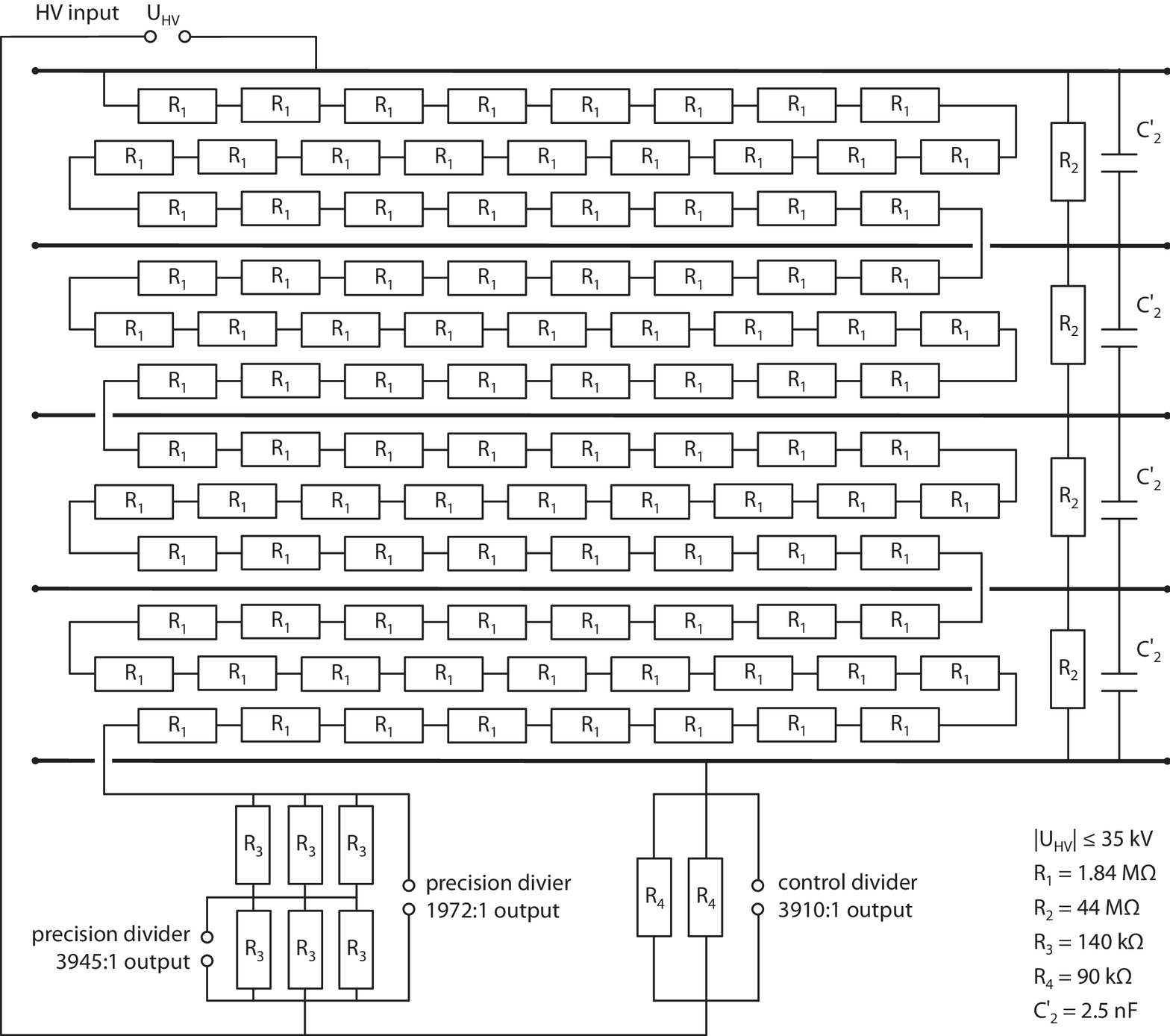}
\caption[Divider circuit overview]{
Circuit overview of the KATRIN precision divider.
Shown are the four main sections with 25 precision resistors of $R_1 = 1.84$~M$\Omega$ each.
The sections are divided by copper electrodes, which are connected by high voltage resistors of $R_2 = 44$~M$\Omega$ and smoothing capacitors of $C_2^\prime = 2.5$~nF.
In the lower part the layout of both precision divider outputs is shown consisting of six precision resistors of $R_3 = 140$~k$\Omega$ each.
The control divider is completed by two standard resistors of $R_4 = 90$~k$\Omega$, which are used to monitor the applied voltage independent of the precision divider.
}
\label{fig:circuit}
\end{figure}

\begin{figure}[t!]
\centering
\includegraphics[width=100mm,angle=-90]{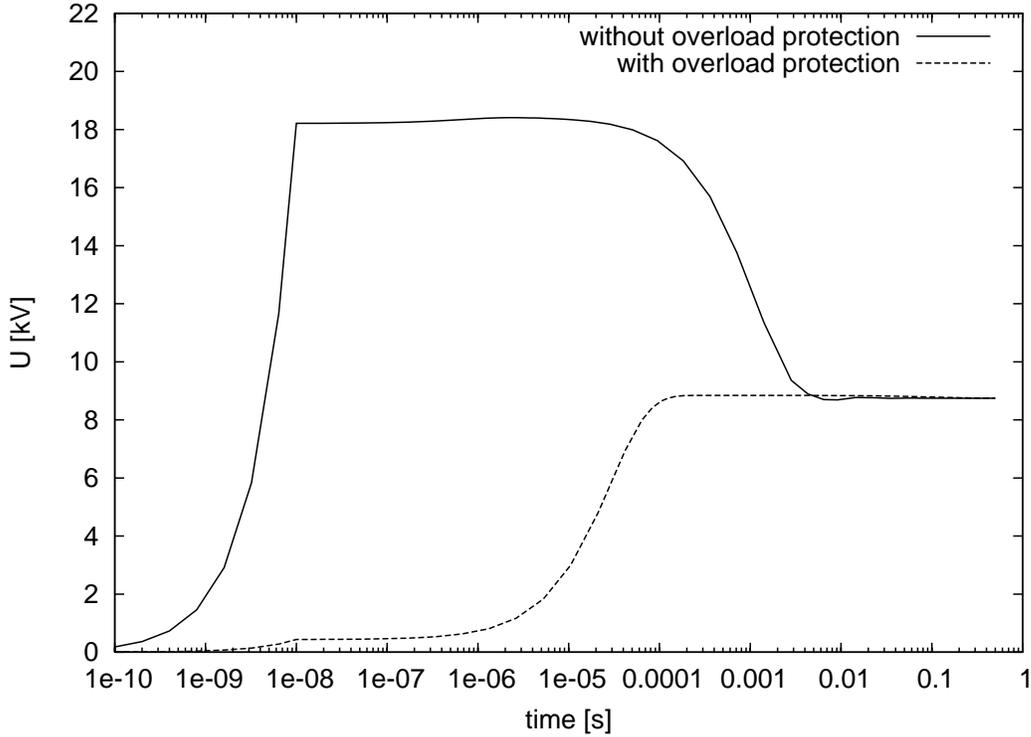}
\caption[PSpice simulation of switching-on response]{
PSpice simulation of the voltage drop over a single electrode section as a function of time.
For this simulation the size of the stray capacitances has been estimated to be $C_1 = 8$ pF between electrode and vessel as well as $C_2 = 14$ pF between two electrodes. 
At $t=0$ a rectangular pulse of 35 kV has been applied.
Without overload protection by additional capacitors between the electrode layers an overload up to 18 kV occurs;
with overload protection by the additional capacitors ($C_2 \rightarrow C_2^\prime = 2.5$ nF) the voltage rises smoothly up to the nominal value of $8.75$ kV for one section.
}
\label{fig:pspice-sim}
\end{figure}

The KATRIN precision high voltage divider is constructed with design parameters similar to those of the PTB standard divider reported in \cite{marx}, but has been adopted to the requirements of the KATRIN experiment of measuring voltages up to 35 kV and uses a completely different precision resistor technology.\\
The inner setup comprises four sections subdivided by five control electrodes made of polished copper and supported by a set of polyoxymethylene (POM) rods, see figures \ref{fig:electrode}, \ref{fig:skizze1}.
Because of its high mechanical and dielectric strength, the thermoplastic POM is used as an insulator and a support structure.  
The high voltage is fed to the top electrode by an appropriate sealed high voltage bushing.
The whole structure is supported by POM rods on the bottom flange of the stainless steel vessel.\\
The high-precision divider consists of 106 precision resistors of Bulk Metal Foil technology (see \sref{screening}).
The layout of the entire divider circuit is shown in \fref{fig:circuit}.
In the four upper sections of the divider 100 resistors of $R_1 = 1.84$ M$\Omega$ are arranged in a helix structure, each section comprises 25 resistors.
The remaining 6 resistors, $R_3 = 140$ k$\Omega$ each, provide two low voltage outputs.
Therefore two groups of three resistors in parallel are arranged subsequent to the 100 precision resistors.
One such group provides a divider ratio of 3945:1, both together provide 1972:1, thus allowing precise measurements in the $10$~V and in the  $20$~V range, as needed by KATRIN.\\
Except for the top electrode, all copper electrodes have a centred bore to fit an acrylic tube, which --- by means of appropriate drillings --- is used to provide a fan-driven flow of temperature stabilized insulation gas ($\mathrm{N_2}$) at each resistor position.
By using a PID microprocessor control unit, a PT100 temperature sensor, suitable heat exchangers, and an external peltier cooling and resistive heating setup, the temperature of the insulation gas is kept stable at $25 ^\circ \mathrm{C} \pm 0.15 ^\circ \mathrm{C}$.
The precision resistors of the high-precision divider are mounted between PTFE rods in order to be kept under a constant gas flow.
The PTFE rods are fixed between the control electrode layers; they are used to prevent or to reduce any leakage and compensating currents between the cylindrical resistor mounts, which are made of nickel-plated brass.
A sketch of one electrode layer is shown in \fref{fig:electrode}.
A POM-insulated feedthrough connects the precision resistor helix of one section with the resistors of the next section.
Each pair of copper electrodes is connected via one HV resistor ($R_2 = 44$~M$\Omega$) and one HV capacitor ($C_2^\prime = 2.5$~nF), forming a capacitive ohmic control divider in parallel to the high-precision divider.
The control divider output consists of two standard resistors of $R_4 = 90$~k$\Omega$ each and does not need to be  calibrated.
In this way the applied voltage can be monitored with low precision and independent of the precision divider.
The HV resistors of the control divider provide a linear voltage distribution in all sections, guaranteeing each precision resistor being placed in an electrostatic potential according to its voltage.
The capacitors of the control divider protect the precision resistors of the high-precision divider from transient overloads, e.g. when the direct high voltage is switched on or off (see \fref{fig:pspice-sim}).
The shape of the outer edge of the copper electrodes has been optimized in order to provide a homogeneous electrostatic field at the mounting position of each measuring resistor.
All parts and structures are designed with edge radii larger than $4$~mm in order to reduce the field strength and to prevent internal discharges.
At the precision resistor mounts the maximum field strength is less than $5$~kV/cm.
Between the $35$~kV copper electrode and the grounded stainless steel vessel the maximum field strength is less than $16.5$~kV/cm.\\
The divider setup is contained in a cylindrical stainless steel vessel filled at standard atmospheric pressure with dry N$_2$ as insulation gas.
Flowing across the precision resistors, the insulation gas is used for temperature stabilization and heat transfer.
Due to its operation in high magnetic fields close to the main MAC-E filter of the KATRIN experiment, the vessel has to be non-magnetic.
It is mounted on top of a mobile 19" rack, which contains the equipment for temperature control and the interface to the KATRIN slow control network (see \fref{fig:skizze1}).

\section{Precision Resistor Selection and Screening Procedure}
\label{screening}

\begin{figure}[t]
\centering
\includegraphics[width=140mm]{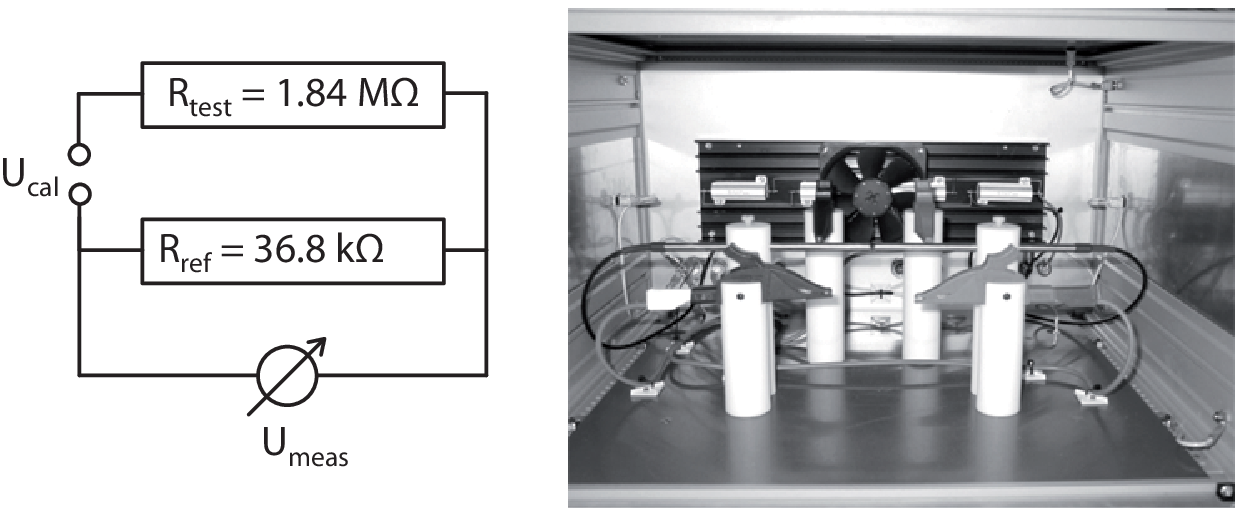}
\caption[warm-up measurement circuit]{
Resistor screening circuit and photo of the view inside the temperature stabilized measurement chamber without resistors installed.
The circuit diagram illustrates how the test resistor is being investigated by monitoring the voltage change at the reference resistor, which has a 50 times lower resistance value.
Due to this only the test resistor is expected to warm up whereas the reference resistor keeps stable in temperature and resistance.
}
\label{fig:meas-circuit}
\end{figure}

\begin{figure}[t!]
\centering
\includegraphics[height=110mm]{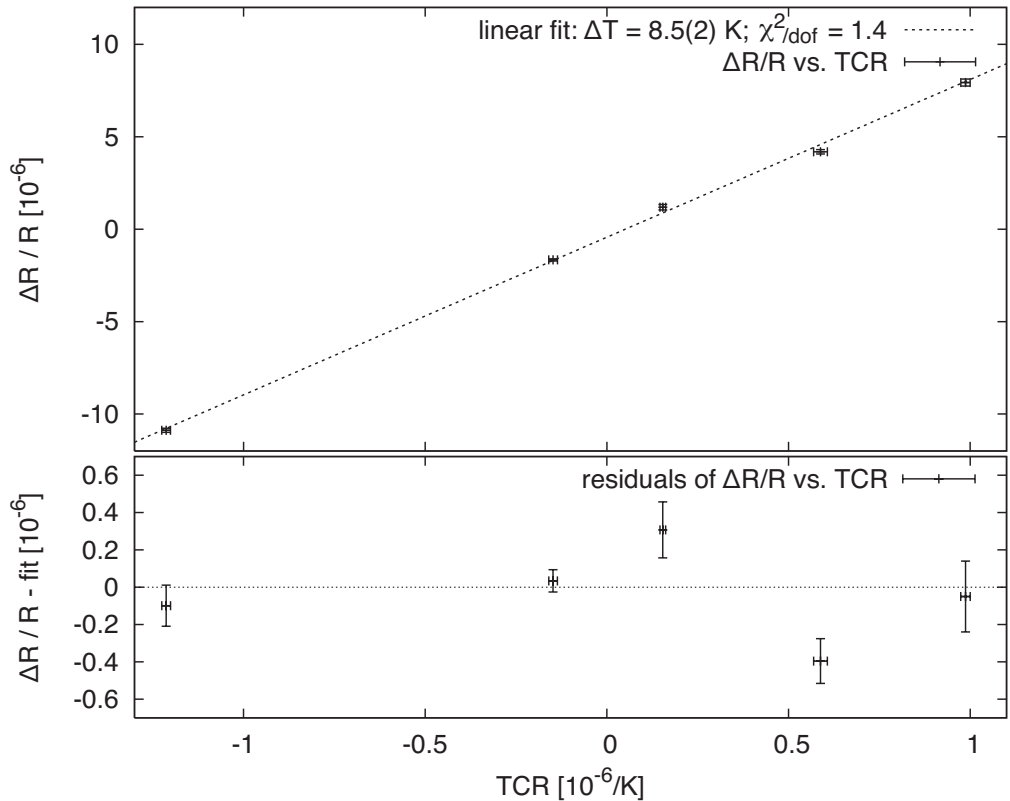}
\caption[Warm-up deviation versus temperature coefficient of resistance]{
Comparison of the warm-up deviation $\Delta R/R$ at a load of $588.2$~V with the independently measured temperature coefficient of resistance for five resistors.
The upper plot shows the linear fit to the data, whereas the differences to the straight line of the linear fit are plotted below. 
}
\label{fig:TCR-vs-warm-up}
\end{figure}

High precision hermetically sealed and oil-filled resistors based on the Bulk Metal Foil\footnote{Bulk Metal Foil is a brand of Vishay Intertechnology, Inc.: www.vishay.com} technology have been chosen to equip the resistor chain of the high-precision divider.
\begin{figure}[t]
\centering
\includegraphics[width=100mm,angle=-90]{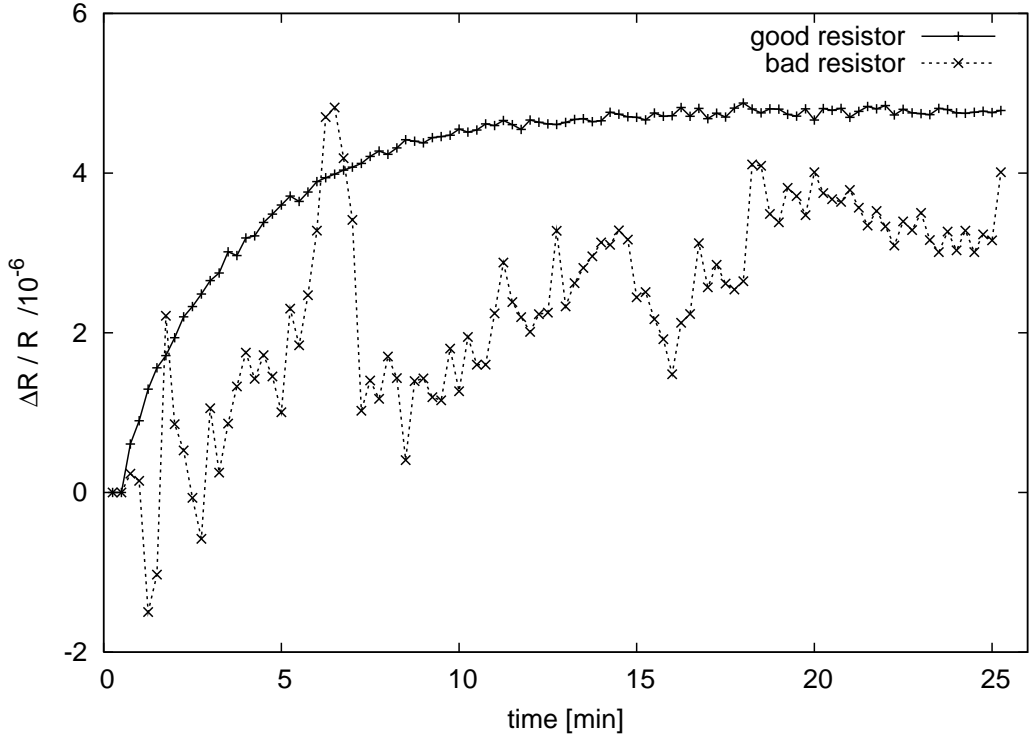}
\caption[Comparison of a good and a bad resistor sample]{
This plot shows the warm-up deviation of two resistors which have been loaded with 588.2~V for 25 minutes.
Both samples keep within their specifications, but one performs as expected and stabilizes after 10 minutes, the other shows an unstable characteristic with no reproducibility.
}
\label{fig:res-compare}
\end{figure}
Those resistors are specified for voltages of up to 600V, they are available with temperature coefficients (TCR) of $\left| \frac{1}{R} \frac{\partial R}{\partial T} \right | < 2 \cdot 10^{-6}~\mathrm{K}^{-1}$ and with high resistance values of up to $R=1.84$ M$\Omega$.
Their design is non inductive and non capacitive and their specified voltage coefficient of resistance (VCR) is $ \left |\frac{1}{R} \frac{\partial R}{\partial U} \right | < 1  \cdot 10^{-7}~\mathrm{V}^{-1}$.
According to the specifications published by the manufacturer the resistors show a long-term stability of $\pm 5 \cdot 10^{-6}$ in one year shelf life\footnote{For three years of shelf life a relative deviation of $\pm 1 \cdot 10^{-5}$ is given.} and $\pm 2  \cdot 10^{-5}$ in load life\footnote{After 2000 h operation at a power of $0.1$ W and a temperature of 60 $^\circ$C.}.
This drift is mainly caused by an ageing effect of the resistor material and is supposed to decrease with time. 
With these values the resistors are about one order of magnitude less precise than the wire-wound resistors used in the PTB standard divider \cite{marx}, which are not anymore commercially available.
The load life stability can be improved by a special pre-ageing procedure, but this procedure has not been applied to the chosen resistors of type VHA-518/11.\\
Under load each resistor shows a characteristic warm-up deviation of the resistance value which is strongly correlated to 
the internal temperature increase and the temperature coefficient of resistance.
This effect is visible even within the specified TCR and VCR values (see \fref{fig:res-compare}) and can be reduced by a careful screening procedure.
Therefore the 100 resistors of 1.84 M$\Omega$ for the high-precision divider have been selected from a lot of 200 resistors by investigating their warm-up deviation.
The screening procedure has been performed within a shielded chamber (see \fref{fig:meas-circuit}) at a stabilized ambient temperature of  $25.0 \pm 0.1 \:^\circ$C.
The measurement circuit that has been used is shown in figure \ref{fig:meas-circuit}; it consists of a calibrated voltage source, one test resistor at a time, the low resistance reference resistor, and a $8\frac{1}{2}$ digit voltmeter (Agilent 3458A).
The calibrated voltage source (Fluke 5720A) supplies the test voltage of $U_\mathrm{cal} = 600$~V with low noise and high stability at the $10^{-6}$ range.
The test resistor ($R_\mathrm{test} = 1.84$ M$\Omega$) and the fifty times smaller reference resistor ($R_\mathrm{ref} = 36.8$ k$\Omega$) create a simple voltage divider.
Due to this, 588.2 V of the applied voltage occurs across the test resistor, whereas 11.8 V occurs across the reference resistor.
With $P_\mathrm{test} =188$~mW the wattage at the test resistor is 50 times higher than at the reference resistor with $P_\mathrm{ref} = 3.8$~mW, hence we can expect that the test resistor will warm up while the effect at the reference resistor is negligible.
When logging the change of the voltage drop at the reference resistor we directly measure the warm-up effect of the test resistor.
In figure \ref{fig:res-compare} it is shown how the initial warm-up deviation stabilizes after about 10 minutes for a typical resistor sample compared to the unstable characteristic of an exceptional bad resistor.
Only resistors whose measured warm-up characteristic matches the following limits are used in the setup:
\begin{itemize}
\item{Temperature coefficient of resistance (TCR): $\left | \frac{1}{R} \frac{\partial R}{\partial T} \right | < 1.2 \cdot 10^{-6}~\mathrm{K}^{-1}$.}
\item{Reproducible and stable operation after $\approx 15$ minutes.}
\item{RMS of final resistance value $< 3 \cdot 10^{-7}$ after warm-up.}
\item{Warm-up deviation of resistance: $\left | \frac{\Delta R}{R} \right | < 10^{-5}$ after 25 minutes.}
\end{itemize} 
\begin{figure}[t!]
\centering
\includegraphics[height=140mm,angle=-90]{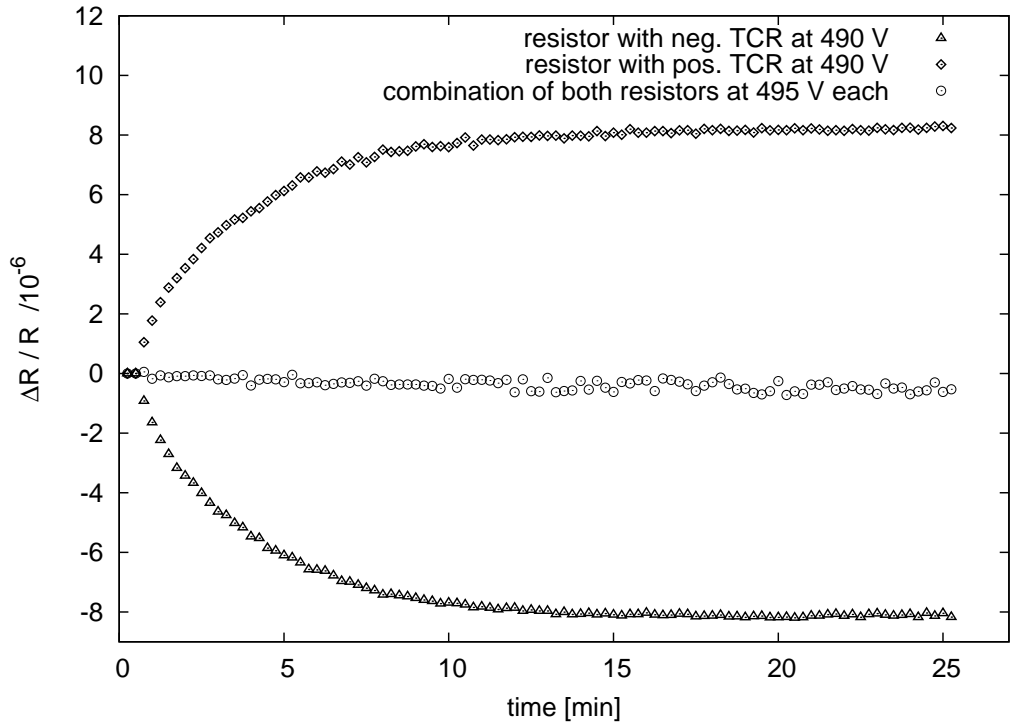}
\caption[Warm-up deviation and compensation]{
Warm-up deviation at a load of 490 V for two single resistors (triangles, diamonds) and at a load of 495 V per resistor for the combination of both (circles).
This set of resistors shows opposite signed TCRs with about the same deviation amplitude.
The combination of both resistors results in an reduced warm-up deviation by more than one order of magnitude.
}
\label{fig:warm-up}
\end{figure}
During operation at the KATRIN experiment the nominal voltage across each resistor will be less than 200~V in Tritium measurement mode when measuring energies around 18.6~keV, with a maximum of 350~V during calibration runs when investigating the \kr\ conversion electrons at energies of up to 32~keV.
In order to investigate the maximal warm-up effect, the maximal rated voltage of 600~V per resistor has been applied to the test circuit, i.e. a load of 588.2~V at the test resistor.
The relative deviation of the resistance value has been monitored for 25 minutes after switching-on.
After the initial warm-up process is finished and the resistor reached thermal equilibrium, the resistance value becomes stable, as shown in \fref{fig:res-compare} for 588.2~V load and in \fref{fig:warm-up} for 490~V load.
By analysing the warm-up deviation against the independently measured TCR value a mean temperature increase of $\Delta T = 8.5 \pm 0.2 \:^\circ$C in each 1.84~M$\Omega$ resistor can be deduced (see \fref{fig:TCR-vs-warm-up}).

Within their specification resistors with positive and negative TCR exist, thus positive and negative warm-up deviations occur.
This property is a result of the Bulk Metal Foil, since it is based on strain gauge technology.
The resistor foil with positive TCR is glued on a ceramic substrate, which shows very low thermal expansion.
The thermal expansion of the resistor foil leads to a rise in resistance, but since it is glued on the substrate a mechanical stress occurs, which leads to a decrease of its resistance.
The low TCR value of the final resistor is achieved by carefully balancing the expansion factors, the glue method, and the substrate material.
Since we investigated the resistors within their specifications, we are sensitive to whether the remaining TCR is positive or negative.

By combining pairs of resistors with identical but different signed warm-up deviations, we were able to reduce the combined warm-up deviation by more than one order of magnitude, see figure \ref{fig:warm-up}.
Finally, pairs or even groups of up to four resistors which in sum show the lowest combined warm-up deviation are used for assembling the resistor chain of the high-precision divider.
The resistors of those pairs and groups are mounted in adjacent positions in order to have identical ambient conditions should there be instabilities in the internal temperature.
Based on the single resistor results, the combined warm-up deviation of the resistance of all 100 selected resistors should be less than $2 \cdot 10^{-8}$ at a load of $588.2$~V per resistor, which is a two orders of magnitude improvement on the average value of a single resistor.\\
The six 140~k$\Omega$ resistors providing the divider ratios are chosen from a lot of 15 resistors; they are matched by reducing their combined temperature coefficient of resistance as well.
No significant warm-up deviation is expected for those resistors because in this part of the divider setup the voltage drop is less than 10~V per resistor.

\section{Investigation of the High-Precision Divider Chain}
\label{invest}

At the laboratory for instrument transformers and high voltage of the PTB Braunschweig we investigated the new high voltage divider in comparison with the PTB reference divider MT100, one of the most precise high voltage dividers in the world, at direct voltages between $-8$ kV and $-32$ kV.
Both dividers (MT100, KATRIN) have been connected to a common precision HV source, thus minimizing the influence of high voltage variations.
The MT100 divider has been upgraded with an additional scale factor of 3334:1 in order to cover the whole range of the precision voltmeter (HP 3458A, 10 V range) when applying voltages below 35 kV.
The MT100 and KATRIN divider output voltages have been monitored by state-of-the-art $8\frac{1}{2}$ digit voltmeters of type HP 3458A and Fluke 8508A, respectively.
A 10~V reference source of type Fluke 732A, calibrated against PTB's Josephson voltage standard \cite{ptb-joseph}, has been used to re-calibrate both digital voltmeters in order to compensate gain and offset deviations before and after each measurement run.
Thermoelectric voltages in particular have to be taken into account when measuring in the 10~V range at $10^{-7}$ relative precision.
In order to reduce any thermoelectric influence, only gold-plated or Cu-Te connectors were used for the measurement chain.
In addition, by encapsulating readout contact pairs of the same material, we achieve identical thermal gradients on both polarities, thus thermoelectric voltages cancel. 
The thermoelectric effect inside the divider housing is expected to be negligible, since the internal temperature of the whole setup is being stabilized.
In this configuration the externally applied voltage $U_\mathrm{HV}$ is related to both divider readings $U_\mathrm{MT100}$, $U_\mathrm{KATRIN}$ and both scale factors  $M_\mathrm{MT100}$, $M_\mathrm{KATRIN}$ according to
\begin{equation}
U_\mathrm{HV}
= U_\mathrm{MT100} \cdot M_\mathrm{MT100}
= U_\mathrm{KATRIN} \cdot M_\mathrm{KATRIN}
\mathrm{.}
\label{eqn:voltage-relation}
\end{equation}
For the scale factor of the KATRIN divider this yields:
\begin{equation}
M_\mathrm{KATRIN}
= \frac{U_\mathrm{MT100}}{U_\mathrm{KATRIN}} \cdot M_\mathrm{MT100}
\label{eqn:scale-factor}
\end{equation}
Series of measurements have been performed repeatedly with identical settings, but independent from one another.
Data sets of these measurements have been combined by compiling a mean value for each time step after applying the voltage.
In order to examine the effect of the scale factor uncertainty on the voltage reading, the relative deviation of the scale factor $M = M_\mathrm{KATRIN}$ based on a reference value $M_0 =  M_\mathrm{0,KATRIN}$ is of interest:
\begin{equation}
\frac{\Delta M}{M_0}
= \frac{M - M_0}{M_0}
=  \frac{U_{\mathrm{MT100}} \cdot U_\mathrm{0,KATRIN}}{U_{\mathrm{KATRIN}} \cdot U_{0,\mathrm{MT100}}} -1 
\label{eqn:rel-dev}
\end{equation}
Depending on the objective of the analysis, the initial, the final, or the average value is used for $M_0$.\\
The main focus of the investigation at PTB was the switching-on deviation, the linearity, and long-term stability of $M_\mathrm{KATRIN}$.
During KATRIN measurement runs the nominal voltage to be monitored is $-18.6$~kV.
In order to take advantage of the whole scale (10~V range) of the precision voltmeter the 1972:1 output is the optimal choice for the given voltage.
The latter configuration has therefore been investigated thoroughly and repeatedly in two calibration campaigns in 2005 and 2006 with a time gap of 13 months.
Measurements performed to check the 3945:1 output show no unanticipated behaviour compared to the 1972:1 output.
In the following subsections the results for the 1972:1 output of the KATRIN divider are presented.
A summary of the results for both scale factors is shown in \tref{tab:summary}.

\subsection{Initial deviation after switching-on}
\label{inidev}

\begin{figure}[t!]
\centering
\includegraphics[height=140mm,angle=-90]{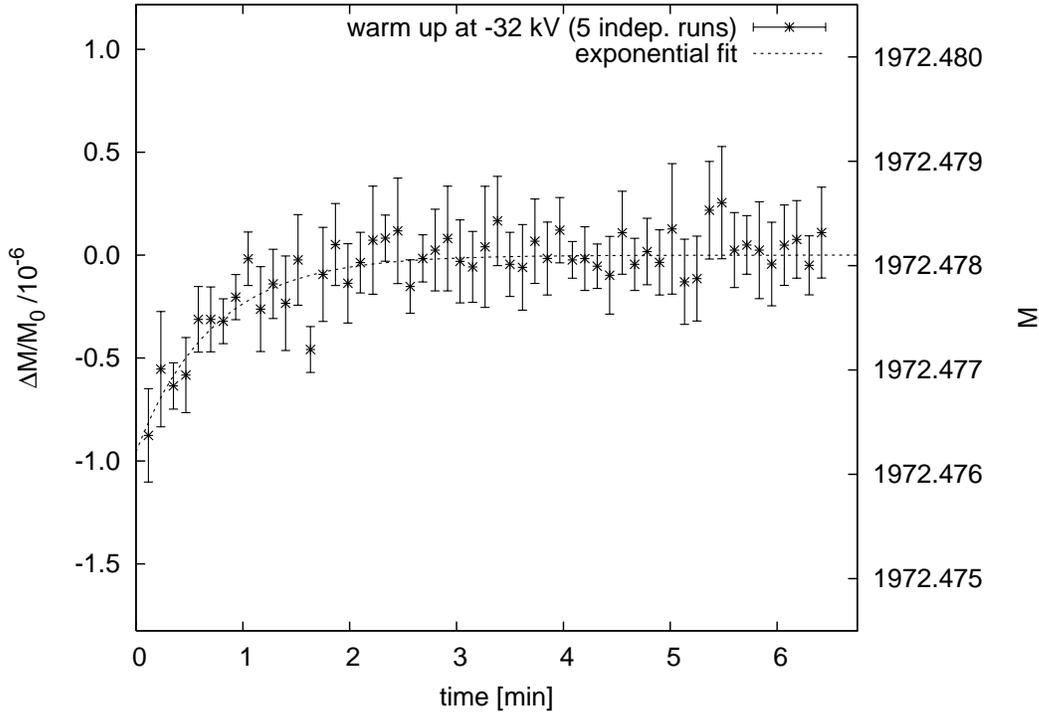}
\caption[Initial deviation at $-32$~kV]{
Deviation of the scale factor M for the 1972:1 output during the first seven minutes after applying high voltage of $-32$~kV.
Plotted data points represent the average per time step of five independent measurements performed under identical conditions.
The error bars denote the standard deviation of the five measurements at each time step.
The exponential fit (dashed line) yields a relative initial warm-up deviation of $1 \cdot 10^{-6}$ during the first two minutes.
The exponential time constant is $0.7$ minutes.
After two minutes the scale factor mean value plus standard deviation keeps stable within a relative deviation of $\pm 5 \cdot 10^{-7}$.
}
\label{fig:inidev}
\end{figure}

Directly after applying the high voltage to the divider, a small warm-up deviation is expected.
This is due to the residual warm-up effect occurring even after the mutual matching has been performed according to the individual warm-up characteristics and the TCRs of the single resistors.
For the MT100 divider it has been demonstrated that this initial warm-up effect is $< 5 \cdot 10^{-8}$ independent of the applied voltage \cite{marx}.
For the KATRIN divider the warm-up deviation has been investigated at $-18$~kV and $-32$~kV for both scale factors.
In all cases, the relative warm-up deviation of the scale factor is about $1 \cdot 10^{-6}$ during the first two minutes of operation; in addition a reproducibility in the $10^{-7}$ range has been observed relative to the absolute scale factor values.\\
An example is shown in \fref{fig:inidev}; here the warm-up effect at the 1972:1 output at $-32$~kV is plotted as a function of time after applying high voltage.
Five independent measurements have been performed, the average of all measurements per time step is plotted as well as statistical error bars.
The remaining fluctuations are dominated by noise from the voltmeter reading.
The exponential fit (dashed line) gives a relative amplitude of about $1 \cdot 10^{-6}$ for the warm-up effect with an exponential time constant of $0.7$ min.
The scale factor stabilizes after two minutes --- after one it is not quite at equilibrium.
This effect has been taken into account in the measurements reported in \sref{linearity}.
The remaining deviation of the mean value plus standard deviation keeps stable within a relative deviation of $\pm 5 \cdot 10^{-7}$.
Since there are five measurements averaged this is a direct demonstration of a reproducibility in the $10^{-7}$ range.\\
It can be summarized that after a short warm-up time the initial scale factor deviation is negligible and stabilizes with a relative reproducibility in the low $10^{-7}$ range, independent of the chosen divider output or operation voltage.
Additionally, since the warm-up effect measured in 2005 has been reproduced in the 2006 measurements, we conclude that it is stable and reproducible for at least 13 months.

\subsection{Linearity and voltage coefficient}
\label{linearity}

\begin{figure}[t!]
\centering
\includegraphics[height=140mm,angle=-90]{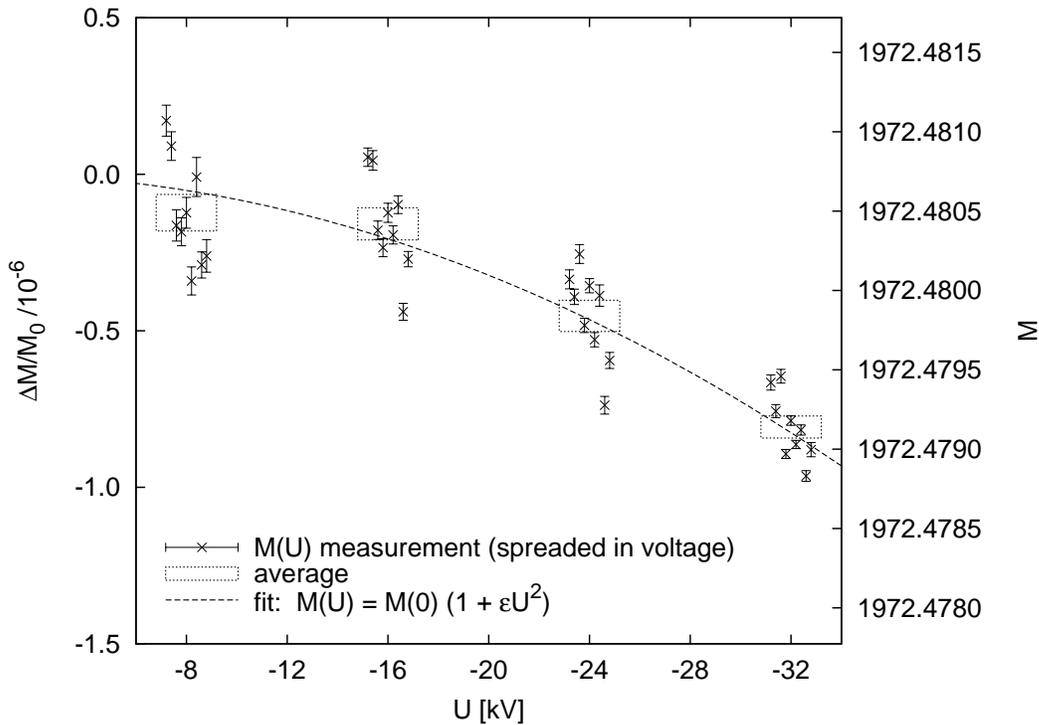}
\caption[Voltage dependence from $-8$~kV to $-32$~kV]{
Voltage dependence of the 1972:1 scale factor for the $-8$~kV to $-32$~kV range.
Measured nine times with identical settings and four hours per voltage measurement time.
The data points show the average scale factor for each single measurement, they have been spread in voltage for visualisation.
The dotted rectangles show the overall average per voltage at the $\Delta M/M_0$ scale.
A quadratic fit (dashed) accounts for the wattage effect and yields a wattage coefficient of $\varepsilon = -8.1(6) \cdot 10^{-10} ~\mathrm{kV}^{-2}$, see text.
}
\label{fig:vcr1972}
\end{figure}

For the KATRIN experiment, calibration and monitoring procedures at voltages between $-17$~kV and $-32$~kV are intended.
A crucial property of a divider is the linearity of the scale factor over the voltage range of interest.\\
Due to the fixed scale factors of the voltage divider of 1972:1 and 3945:1, which have been optimized for scaling down voltages of about $-18$~kV to the $10$~V range of precision digital voltmeters, the systematic uncertainty of the low voltage measurement increases significantly at primary voltages $|U| \leq 10$~kV, e.~g. when reading low voltages of less than $\pm 5$~V the relative uncertainty of the reading of high precision voltmeters exceeds $5 \cdot 10^{-7}$.
Nevertheless, the linearity of the KATRIN divider has been investigated at voltages between $-8$~kV to $-32$~kV with $2$~kV steps.
Hence, when setting the high voltage to $-8$~kV, low voltages of about $-4$~V have to be read at the 1972:1 output of the new divider and about $-2.4$~V at the 3334:1 output of the MT100 divider, respectively.\\
Each of the voltage steps has been applied for two minutes, whereas only the data of the second minute has been evaluated due to the initial warm-up effect which is reported in section \ref{inidev}.
\begin{figure}[t!]
\centering
\includegraphics[height=140mm,angle=-90]{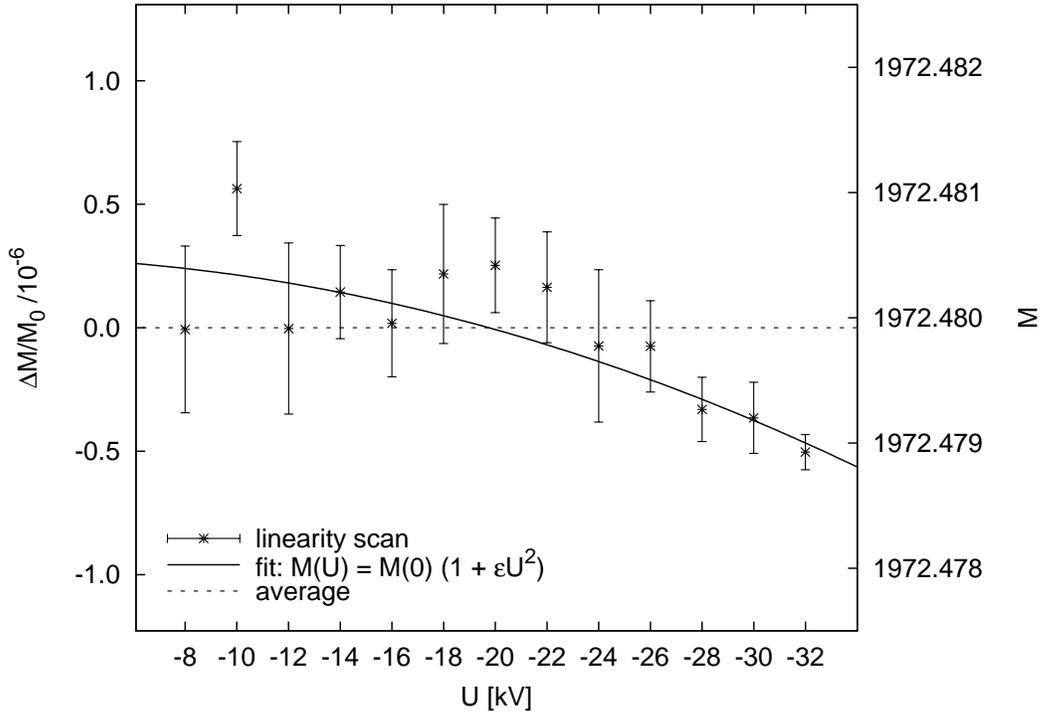}
\caption[Linearity between $-8$~kV and $-32$~kV]{
Linearity of the KATRIN divider as evaluated from five independent measurement runs.
Plotted is the averaged scale factor as a function of the applied voltage.
The error bars denote the standard deviation of the five measurements at each voltage setting. 
As described in the text, the uncertainty of the voltage reading increases at $|U| \leq 10$~kV.
At voltages $|U| > 10$~kV and close to the Tritium endpoint energy, the linearity of the divider is in the $10^{-7}$ range.
The scale factor keeps stable within a $5 \cdot 10^{-7}$ interval around the average for $12\:\mathrm{kV} \leq |U| \leq 26\:\mathrm{kV}$.
The quadratic fit $M(U)$ accounts for the wattage effect.  
}
\label{fig:linearity}
\end{figure}
The combined result of all five independent linearity investigations is shown in figure \ref{fig:linearity}.
Each plotted scale factor incorporates the average of all measurements for the same voltage setting. 
As expected, the spread and the uncertainty increase at voltages $|U| \leq 10$~kV.
At voltages $|U| > 10$~kV and especially at a wide band around the Tritium endpoint energy, the linearity of the divider is in the $10^{-7}$ range.
Here, the spread is less than $3 \cdot 10^{-7}$, which has been reproduced in all five independent measurement runs.\\
The slope at the high voltage end indicates a voltage dependence that has to be investigated in more detail.
Since the wattage increases quadratically with the applied voltage, the self heating of each resistor should increase quadratically as well.
The fit function 
\begin{equation}
M(U) = M(0) \cdot \left ( 1 + \varepsilon U^2 \right )
\label{eqn:wattage-fit}
\end{equation}
accounts for this case and yields a zero voltage scale factor $M(0)$ and the wattage coefficient $\varepsilon$.
In \fref{fig:linearity} the fit of \eref{eqn:wattage-fit} to the data delivers $M(0) = 1972.4805(2)$ and $\varepsilon = -7.4(1.2) \cdot 10^{-10} ~\mathrm{kV}^{-2}$ with a reduced $\chi^2$ of $0.8$.
The result of this evaluation has large uncertainties since the measurement time and statistics of the linearity scan is limited.\\
Therefore, nine independent long-term measurements with a voltage stepping of $-8$~kV, $-16$~kV, $-24$~kV, and $-32$~kV have been performed. 
In order to assure stable measurement conditions and higher statistics, each voltage setting has been applied and monitored for four hours.
\Fref{fig:vcr1972} shows the result of all measurements at the 1972:1 output by plotting the evaluated average per measurement and voltage and the overall average per voltage.
In order to account for the wattage effect a quadratic fit \eref{eqn:wattage-fit} is applied. 
It yields a zero voltage scale factor of $M(0) = 1972.4807(1)$ and a wattage coefficient of $\varepsilon = -8.1(6) \cdot 10^{-10} ~\mathrm{kV}^{-2}$ which is in agreement with the linearity measurement done earlier.
In the voltage range of $-8$~kV to $-32$~kV this leads --- if unaccounted --- to a relative deviation of $0.77(9) \cdot 10^{-6}$ due to the voltage dependence and the wattage effect accordingly.
Taking into account the whole voltage range from zero to $-35$~kV the relative deviation increases to $0.99(10) \cdot 10^{-6}$, which is still well within the requirements of the KATRIN experiment.\\
For the 3945:1 output one gets a similar result with  $M(0) = 3944.9612(1)$ and  $\varepsilon = -7.5(4) \cdot 10^{-10} ~\mathrm{kV}^{-2}$, which corresponds to a relative deviation of  $0.72(6) \cdot 10^{-6}$ in total for the voltage range $-8$~kV to $-32$~kV.
Accordingly, the relative deviation over the whole voltage range from zero to $-35$~kV is  $0.91(7) \cdot 10^{-6}$, which is still well within the requirements of the KATRIN experiment as well.\\
The voltage dependence results for both scale factors agree within their uncertainty and since this result has been measured in 2005 and reproduced in 2006 we can conclude that the effect is stable and that the scale factor deviations due to voltage dependence and wattage effect are well below the uncertainty limit for the KATRIN experiment.
Moreover, based on these results linearized voltage coefficients of both scale factors can be derived from
\begin{equation}
\frac{1}{M} \frac{\partial M(U^\prime)}{\partial U^\prime} = 2  \varepsilon U^\prime
\label{eqn:linear-vcr}
\end{equation}
for a certain retarding potential $U^\prime$.
In the case of KATRIN the most commonly monitored voltage will be $U^\prime = -18.6$~kV which corresponds to the energy filter setting at the tritium endpoint.
At this voltage \eref{eqn:linear-vcr} yields a linearized voltage coefficient of $\frac{1}{M} \frac{\partial M(U^\prime)}{\partial U^\prime} = -3.0(2) \cdot 10^{-8}$~kV$^{-1}$ for the 1972:1 output and $\frac{1}{M} \frac{\partial M(U^\prime)}{\partial U^\prime} = -2.8(2) \cdot 10^{-8}$~kV$^{-1}$ for the 3945:1 output respectively.
This shows that the voltage dependence is negligible in the tritium endpoint investigations for the neutrino mass determination.

\subsection{Temperature dependence}
\label{tempdep}

Although the new divider is equipped with a sophisticated temperature control, showing a maximum fluctuation of only $0.1$~K, the temperature dependence of both outputs has been investigated at $25\:^\circ$C and at $30\:^\circ$C.
During this test the environmental temperature in the lab was stable at $(21 \pm 1 ) ^\circ$C.
In the beginning, the whole setup has been operated at a stable temperature control set-point of $25\:^\circ$C.
After several hours of stable operation the temperature has been increased to $30\:^\circ$C by adjusting the temperature control set-point.
After the scale factor reading had settled and the divider has been running under stable conditions again, the high temperature scale factor value was measured for several hours.
Finally, the temperature control set-point has been set back to the initial value of $25\:^\circ$C in order to reproduce the initial scale factor value.
The relative temperature deviation found for the 1972:1 output is $\frac{1}{M} \frac{\partial M}{\partial T} = -8.1(6) \cdot 10^{-8} ~\mathrm{K}^{-1}$.
The result for the 3945:1 output is $\frac{1}{M} \frac{\partial M}{\partial T} = 1.71(73) \cdot 10^{-7} \mathrm{K}^{-1}$.
Both values are far lower than the temperature coefficients of any commercial high voltage divider, which is again a measure for the quality achieved by the screening and matching procedure of the precision resistors.\\
In order to check the performance of the internal temperature stabilization, the divider has been operated at environmental temperatures between $20\:^\circ$C and $30\:^\circ$C with a temperature control set-point of $25\:^\circ$C.
In this test it has been demonstrated that the internal temperature stabilization is able to maintain stable conditions at environmental temperatures of up to $27\:^\circ$C.
Since we expect temperatures of  $(21 \pm 3 ) ^\circ$C in the laboratory environment of KATRIN, we can expect a stable operation of the divider without any temperature dependence.\\
We conclude that with respect to the fluctuations of the temperature control, the temperature dependence of both outputs as well as of the complete set-up under laboratory conditions is negligible.
On the other hand the difference in sign and value between both scale factors indicates the technical limit of the matching procedure of the precision resistors used in the low voltage tap of the divider.

\subsection{Medium-term measurements}
\label{medium-term}

\begin{figure}[t!]
\centering
\includegraphics[height=140mm,angle=-90]{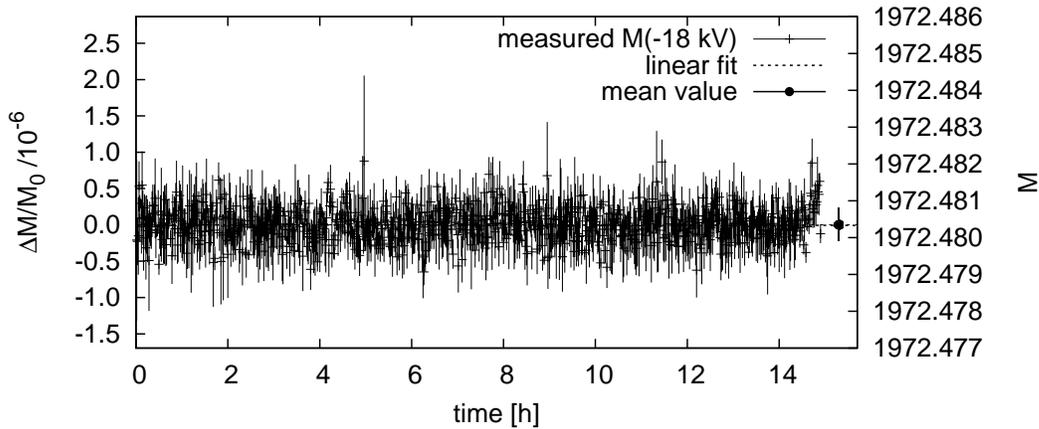}
\caption[Medium-term measurement at $-18$~kV]{
Medium-term measurement at $-18$~kV.
Measurement time is 15 hours.
Plotted is the compiled average of three independent medium-term measurements with
identical parameters.
The error bars (vertical lines) represent the standard deviation of all three measurements per time-step.  
The mean value (separated point at the right) of all scale factor measurements is $M_0 = 1972.4804(5)$ with a relative standard deviation of $\sigma = 0.25 \cdot 10^{-6}$.
A linear fit yields a negligible slope of $-0.011(43) \cdot 10^{-6}$ per day ($\chi^2/dof = 0.94$).
}
\label{fig:long-term-1972}
\end{figure}

In order to investigate the scale factor stability during medium-term measurement runs, a series of three 15 hours long overnight measurements has been performed at a voltage of $-18$~kV and at stable environmental conditions.
All measurements have been averaged time-step by time-step, the result is plotted in \fref{fig:long-term-1972}.
In order to investigate the presence of a time dependent drift of the scale factor, a linear fit has been applied yielding a slope of  $-0.011(43) \cdot 10^{-6}$ per day with $\chi^2/dof = 0.94$.
The average scale factor of this data-set is $M_0 = 1972.4804(5)$ with a relative standard deviation of $\sigma = 0.25 \cdot 10^{-6}$.
Hence, no significant drift can be observed as long as the environmental conditions are stable.
Changing the room temperature conditions results in a small deviation, see the deviation at $t > 14$~h when the door of the laboratory has been opened in the morning.
This is no effect of the voltage divider, but of auxiliary equipment like the digital voltmeters.  
With this result the one day stability of the divider can be rated as significantly lower than the {KATRIN} stability limit.
Nevertheless, no statement on the scale factor drift or an extrapolation over longer time periods is reasonable, based on this kind of medium-term measurements.

\subsection{Long-term stability}
\label{long-term-stab}

As pointed out in section \ref{medium-term}, it was not feasible to derive a significant long-term deviation by a single measurement nor by the entire measurement campaign at PTB. 
In order to make a reasonable statement on the long-term stability a longer time interval is needed, therefore the absolute scale factor results of both measurement campaigns in October 2005 and in November 2006 have been compared with each other.
The absolute scale factor values of both measurement campaigns have been derived by averaging all measurement runs with at least two hours of measurement time and independent of the voltage setting.
The resulting absolute scale factors of each measurement campaign are listed in \tref{tab:summary}.
Between the 2005 and the 2006 measurements, several parts of the precision low voltage measurement equipment at PTB have been improved further, resulting in lower uncertainties of both scale factor results of 2006.
Comparing the values of both years\footnote{The PTB standard divider MT100 is being re-calibrated frequently more than twice per year by well established PTB methods, which results in a well traceable relative long-term stability of $2 \cdot 10^{-6}$/year.} we derive a relative scale factor drift of $6.0 \cdot 10^{-7}$/month for the 1972:1 output and  $5.6 \cdot 10^{-7}$/month for the 3945:1 output.
These values are supported by a comparison of measurements of the K-32 conversion electrons of \kr\ in 2006 and 2007 \cite{phd_ostrick2008}.
It is obvious that those small values are not detectable in a single day measurement nor over several days of measurements.
A possible explanation for this drift is the ageing effect of the precision resistors, which have not been pre-aged during the production process.
According to their specifications (see \sref{screening}) this drift will decrease with time, but as of now it requires a re-calibration of the scale factors on a frequent basis at least twice per year in order to keep the relative uncertainty within the  KATRIN limit of  $3.3 \cdot 10^{-6}$.
In future voltage dividers it is strongly recommended to use pre-aged resistors.

\subsection{Estimation of uncertainty}

\begin{table}[t]
\begin{center}
\caption[Uncertainty budget (short version) for the KATRIN HV divider at $-18.6$~kV]{Uncertainty budget (short version) for the KATRIN HV divider at $-18.6$~kV. Listed are the absolute uncertainty contributions for the scale factor determination at the 1972:1 output. The combined standard uncertainty of $1.99 \cdot 10^{-3}$ corresponds to a relative uncertainty of $1.0 \cdot 10^{-6}$.}
\label{tab:uncertainty}
\vspace{2mm}
\lineup
\begin{tabular}{lcc}
\br
Source of Uncertainty	&\multicolumn{2}{l}{Uncertainty contribution}\\
					&absolute		&relative at\\
					&$\times 10^{-3}$	&1972:1 output\\
\mr
PTB standard divider MT100 (3334.65086:1 output)	&$1.77$   &$8.97 \cdot 10^{-7}$\\[1.5ex]
Spread of the ratio of the divider output &   &\\
voltages during a series of measurements	&0.67   &$3.40 \cdot 10^{-7}$\\
 in 7 days at $T_\mathrm{lab} = (22.0 \pm 0.2) ^\circ$C       &   &\\[1.5ex]
DVM of the PTB standard divider	&$0.34$   &$1.72 \cdot 10^{-7}$\\[1.5ex]
DVM of the KATRIN divider	         &$0.23$   &$1.17 \cdot 10^{-7}$\\[1.5ex]
Short-term stability of the voltage source &   &\\
during the instants of measurement with &$0.23$   &$1.17 \cdot 10^{-7}$\\
the two DVMs (ripple $< 1 \cdot 10^{-5}$)	&   &\\[1.5ex]
PTB standard divider drift (whole meas. phase)  	&$0.40$   &$2.03 \cdot 10^{-7}$\\[1.5ex]
Combined standard uncertainty	&$1.99$   &$1.01 \cdot 10^{-6}$\\
\br
\end{tabular}
\end{center}
\end{table}

The uncertainty of the scale factors of the KATRIN HV divider was estimated according to the ISO Guide \cite{ISO-uncetainty}.
Following the concept of the ISO Guide, a measurement yields only an approximate value of the measurand and the uncertainty of the measurement characterizes the interval that encompasses a large fraction of all probable values (coverage probability).
In general, the value and uncertainty of a measurand depends on a number of input quantities, and their functional relationship is expressed by the model function.
\Tref{tab:uncertainty} shows an example of estimating the uncertainty for the 1972:1 output of the KATRIN HV divider at $-18.6$~kV, i.e. the voltage at the endpoint of the tritium \bspec.
The model equation for the divider scale factor and its uncertainty determined by comparison with the PTB standard divider is based on \eref{eqn:scale-factor}.
In addition to the uncertainty of the PTB standard divider \cite{marx} five uncertainty contributions have been considered.
The uncertainty was calculated under the assumption that there is no correlation among the input quantities, using a commercial software \cite{GUM}.
The scale factor of the KATRIN HV divider and its expanded uncertainty at $-18.6$~kV is:
\begin{eqnarray}
\label{eqn:scale-factor-18.6}
M_{\mathrm{KATRIN},\,-18.6\,\mathrm{kV}} &=& 1972.4801 \pm 0.0040  \\
 &=& 1972.4801(1 \pm 2 \cdot 10^{-6}) \nonumber
\end{eqnarray}
for a coverage probability of approximately $95 \%$~$(\mathrm{k} = 2)$\footnote{The uncertainty expansion factor $\mathrm{k} = 2$ is equivalent to a $2\sigma$ uncertainty statement.}.
The result is valid for a warm-up time of the KATRIN HV divider of at least 5 min.
The long-term stability of the KATRIN HV divider has not been taken into consideration because it will be corrected according to \sref{long-term-stab}.\\
A summary of all investigated divider properties is shown in \tref{tab:summary}.
The scale factor values given there are averaged over the whole voltage range in order to cover the overall divider performance, independent of the applied voltage.
The uncertainty contributions of \tref{tab:uncertainty} are not contained in \tref{tab:summary}.
Nevertheless, the averaged scale factor for the 1972:1 output (Nov. 2006) is in agreement with \eref{eqn:scale-factor-18.6}.
This illustrates that the error calculation of \eref{eqn:scale-factor-18.6} is correct.
Applying the long-term stability corrections as described in \sref{long-term-stab}, we conclude that the divider fulfils the requirement on the stability of the KATRIN HV monitoring system.

\begin{table}[t]
\begin{center}
\caption[Summary of calibration against PTB MT100]{Summary of the calibration parameters which have been deduced from all investigations of the new KATRIN divider against the MT100 divider of PTB.
The absolute scale factor values of both measurement campaigns have been derived by averaging all measurement runs with at least two hours of measurement time and independent of the voltage setting.
The given uncertainties are purely statistical, for the absolute calibration including systematics please refer to \tref{tab:uncertainty}.}
\label{tab:summary}
\vspace{2mm}
\lineup
\begin{tabular}{lll}
\br
Parameter		&\m1972:1 output		&\m3945:1 output\\
\mr
\multicolumn{3}{c}{Results of Oct. 2005 measurements} \\
Scale factor $M_\mathrm{KATRIN,\,Oct. 2005}$		&\m1972.4645(11)\,:\,1		&\m3944.9305(21)\,:\,1\\[0.5ex]
Relative standard deviation	&$\m5.5 \cdot 10^{-7}$			&$\m5.2 \cdot 10^{-7}$\\[1.5ex]
\multicolumn{3}{c}{Results of Nov. 2006 measurements} \\
Scale factor $M_\mathrm{KATRIN,\,Nov. 2006}$		&\m1972.48016(61)\,:\,1		&\m3944.9597(14)\,:\,1\\[0.5ex]
Relative standard deviation	&$\m3.1 \cdot 10^{-7}$			&$\m3.5 \cdot 10^{-7}$\\[2ex]
\multicolumn{3}{c}{General characteristics} \\
Temperature dependence	&$-8.1(6) \cdot 10^{-8}$/K	&$\m1.7(7) \cdot 10^{-7}$/K\\[0.5ex]
Temperature stability	&\multicolumn{2}{c}{$\pm 0.1$ K}\\[2ex]
Linearized voltage dep. at $-18.6$~kV     &$-3.0(2) \cdot 10^{-8}$/kV   &$-2.8(2) \cdot 10^{-8}$/kV \\[0.5ex]
Wattage coefficient $\varepsilon$	    &$-8.1(6) \cdot 10^{-10}$/kV$^2$   &$-7.5(4) \cdot 10^{-10}$/kV$^2$\\[0.5ex]
Relative shift due to $\varepsilon$ for $24$~kV    &$\m0.77(9) \cdot 10^{-6}$      &$\m0.72(6) \cdot 10^{-6}$   \\[0.5ex]
Voltage range 	&\multicolumn{2}{c}{$-8$\,kV to $-32$\,kV}\\[2ex]
Warm-up deviation (see \fref{fig:inidev}) &\multicolumn{2}{c}{$1 \cdot 10^{-6}$}\\[0.5ex]
Warm-up time constant (see \fref{fig:inidev})  &\multicolumn{2}{c}{$0.7$ min}\\[2ex]
Long-term stability		&$\m6.0 \cdot 10^{-7}$/month	&$\m5.6 \cdot 10^{-7}$/month\\
\br
\end{tabular}
\end{center}
\end{table}

\section{Conclusion and outlook}
\label{conclusion}

Summarizing the investigation of the KATRIN precision high voltage divider compared with the reference divider of PTB Braunschweig, it is obvious that the resistor screening and matching according to the warm-up deviation was successful.
It was possible to improve the stability of the combined set-up by more than one order of magnitude compared to the properties of a single resistor.
In addition the temperature dependence of the divider is one order of magnitude lower than that of a single resistor.
Due to the temperature stabilization of the whole set-up and its independence of the lab temperature in a certain range, the net temperature dependence of the voltage reading is negligible.
The warm-up deviation of the KATRIN divider has been reduced to about $1 \cdot 10^{-6}$ relative to the scale factor and remained stable over all measurements in 2005 and 2006.
The voltage and wattage dependence of the scale factors over the specified voltage range has even been reduced to a relative deviation of less than $1 \cdot 10^{-6}$ if not corrected, otherwise it is even smaller.
Moreover, at the voltage setting for tritium endpoint investigations at $-18.6$~kV the linearized voltage coefficient is negligible.
With these properties the new divider fulfils the requirement of the {KATRIN} experiment of a relative stability of less than $3.3 \cdot 10^{-6}$.\\
Especially during one cycle of data taking, i.e. one cycle of the source of 60 days, the precision and stability is better than specified.
But since a long-term drift of the setup of $6.0 \cdot 10^{-7}/\mathrm{month}$ (1972:1 output) has been observed, it is recommended to re-calibrate the scale factors twice a year or to perform an on-line calibration during data taking in order to compare different data taking cycles with each other.
The latter will be done with the KATRIN monitor spectrometer in parallel with the KATRIN main beam-line.
While measuring the tritium spectrum at the main beam-line, it is intended to monitor a mono-energetic conversion electron source based on the isomeric state of the isotope \kr\ at the monitor spectrometer.
In this way it is possible to monitor the scale factor drift of the divider and to re-calibrate it frequently.
However, independent re-calibration investigations will be performed at PTB as well.
For redundancy reasons and in order to prevent down-time of the KATRIN measurement during the calibration at PTB, a second divider is being built.
Several improvements applied to its design, especially the use of pre-aged and optimized precision resistors of the same brand, promise further improvement in overall and in long-term stability.

\ack
\label{acknow}

We thank the Physikalisch-Technische Bundesanstalt (PTB) and especially the head of the working group for instrument transformers and high voltage, Dr.-Ing.~K. Schon, for his support and  the opportunity to investigate the new divider there in two extensive measurement campaigns.
This work is supported by BMBF under grant No. 05CK5MA/0.

\end{document}